\documentclass[11pt,twoside]{article}

\usepackage{asp2006_hotcool}
\usepackage{epsf}
\usepackage{lscape}

\begin{document}

\title{A Census of Massive Stars Across the Hertzsprung-Russell Diagram of Nearby Galaxies: What We Know and What We Don't}
\author{Philip Massey}
\affil{Lowell Observatory, 1400 W. Mars Hill Road, Flagstaff, AZ 86001 USA}

\begin{abstract}

When we look at a nearby galaxy, we see a mixture of foreground stars and bona fide extragalactic stars.
 I will describe what we need to do to get meaningful statistics on the massive star populations across the H-R diagram.
Such a census provides the means of a very powerful test of massive star evolutionary theory.
\end{abstract}


\section{Introduction}   

In this review, I will briefly discuss what we do (and don't!) know about the population of massive
stars across the H-R diagram (HRD) of nearby, Local Group galaxies.   If we are accurate census
takers, then we can use the numbers of differing types of massive stars (O-type, Luminous Blue Variable stars, F and G supergiants,
red supergiants, Wolf-Rayet stars) to provide a sanity check on the predictions of stellar evolution theory,
and in particular to see if the relative numbers change as a function of metallicity in the way
that they should.  For this, we use those galaxies of the Local
Group which are currently forming massive stars.
These span a range of a factor of 20 or so in metallicity
as measured by the oxygen content of their HII regions (Massey 2003), and
thus allow a more powerful test than merely that offered by comparing the content of (say) the SMC
to the Milky Way, where the metallicities differ only by a factor of 4 or so.

We expect that metallicity $z$ matters, primarily as the mass loss rates scale as something like
$z^{0.7}$  (Vink et al.\ 2001) during the hydrogen-burning main-sequence phase, and the amount of mass
loss during this phase plays a dominant role in the evolution of massive stars.  Our modern
understanding of this traces back to Peter Conti, who  proposed  (Conti 1976)
that Wolf-Rayet stars were the
result of mass loss.  As a massive star evolves, it loses mass, and eventually reveals He and N
(the products of CNO burning) at the surface, at which point the star is spectroscopically
identified as a WN star.  Further mass-loss eventually reveals C and O, the products of He-burning,
and the star is called a WC.  Conti and collaborators eventually  argued that the metallicity dependence of the mass loss rates
were responsible for the very different relative number of WC- and WN-type Wolf-Rayets seen in
external galaxies (Vanbeveren \& Conti 1980; see also Massey \& Johnson 1998 and Massey 2003).
We will discuss this more extensively below, but this is the sort of test such a census allows.

This is a good moment to conduct such tests.  On the one hand, our Local Group Galaxies survey
(Massey et al.\ 2006, 2007a, 2007b) has used the KPNO and CTIO 4-m telescopes to perform photometry
of stars throughout the star-forming galaxies of the Local Group.  On the other hand, evolutionary theory
now includes the important effects of rotation at various metallicities (Meynet \& Maeder 2003, 2005), and thanks
to the generosity of our colleagues in Geneva the models are freely available.  Still, nothing worth doing is easy, and
we will find that while there are a lot of things that we {\it do} know now, there is still quite a bit that we don't.
There is still plenty of fun to be had.

\section{Massive Star Evolution: An Observer's View}

I'd like to start by sharing my (limited) understanding of what stellar evolutionary theory predicts,
using as our baseline the Geneva evolutionary tracks for $z=0.020$ (solar) of Meynet \& Maeder (2003) and
an initial rotation of 300 km~s$^{-1}$.  This way when we talk about a census of stars across the HRD we will all have
in mind the same mental picture of where these stars are coming from.

First, let us consider a star of very high mass, 120$M_\odot$.  In Fig.~1(a) we plot its path in the HRD,
with the main-sequnce (MS),  Luminous Blue Variable (LBV), and Wolf-Rayet (WR)
phases (WN, WC) indicated.    At the beginning the star is spectroscopically an O3~V, but  the
star then enters the WN phase while still a main-sequence (core-H burning) object.

\begin{figure}[!ht]
\plottwo{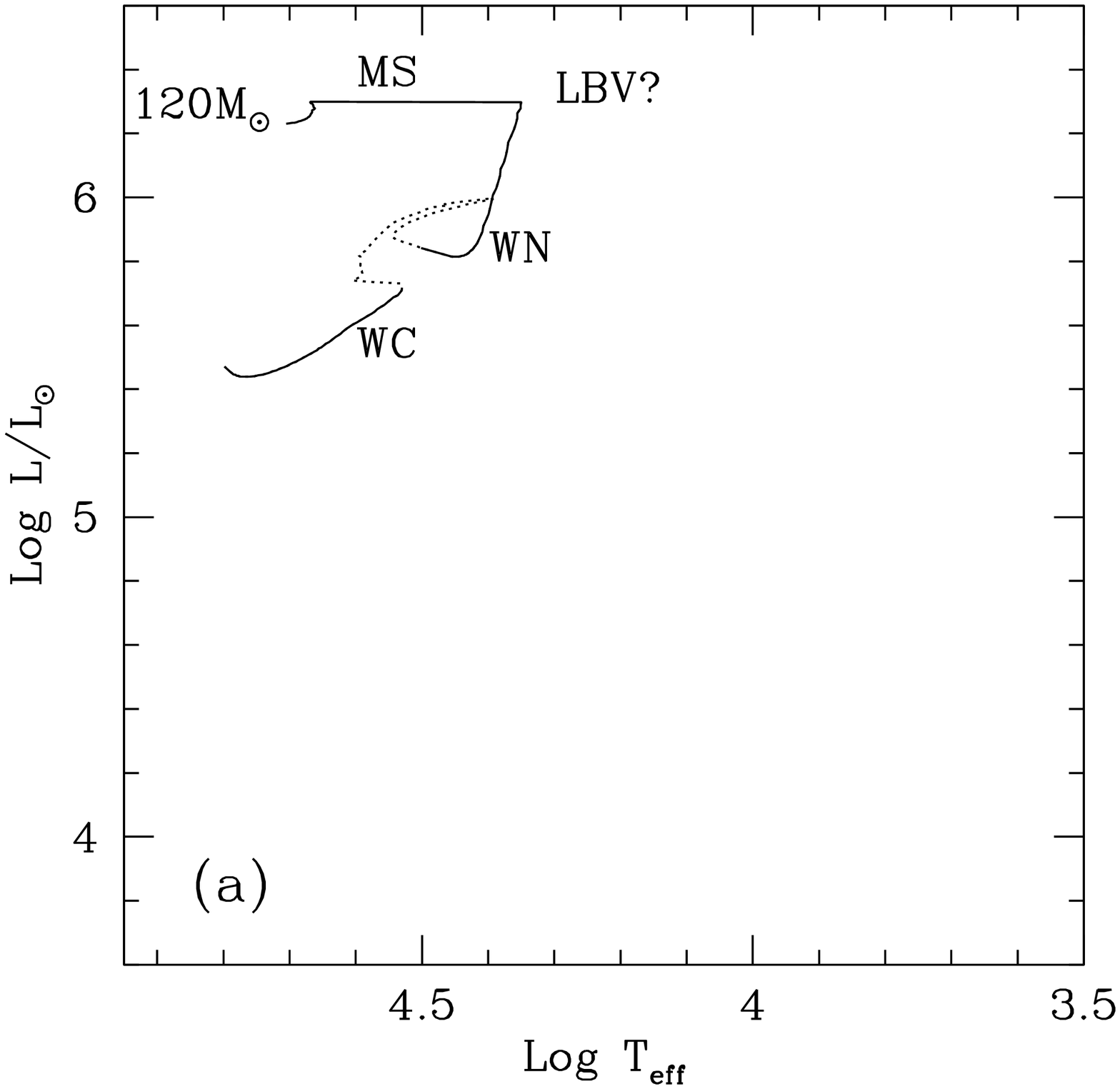}{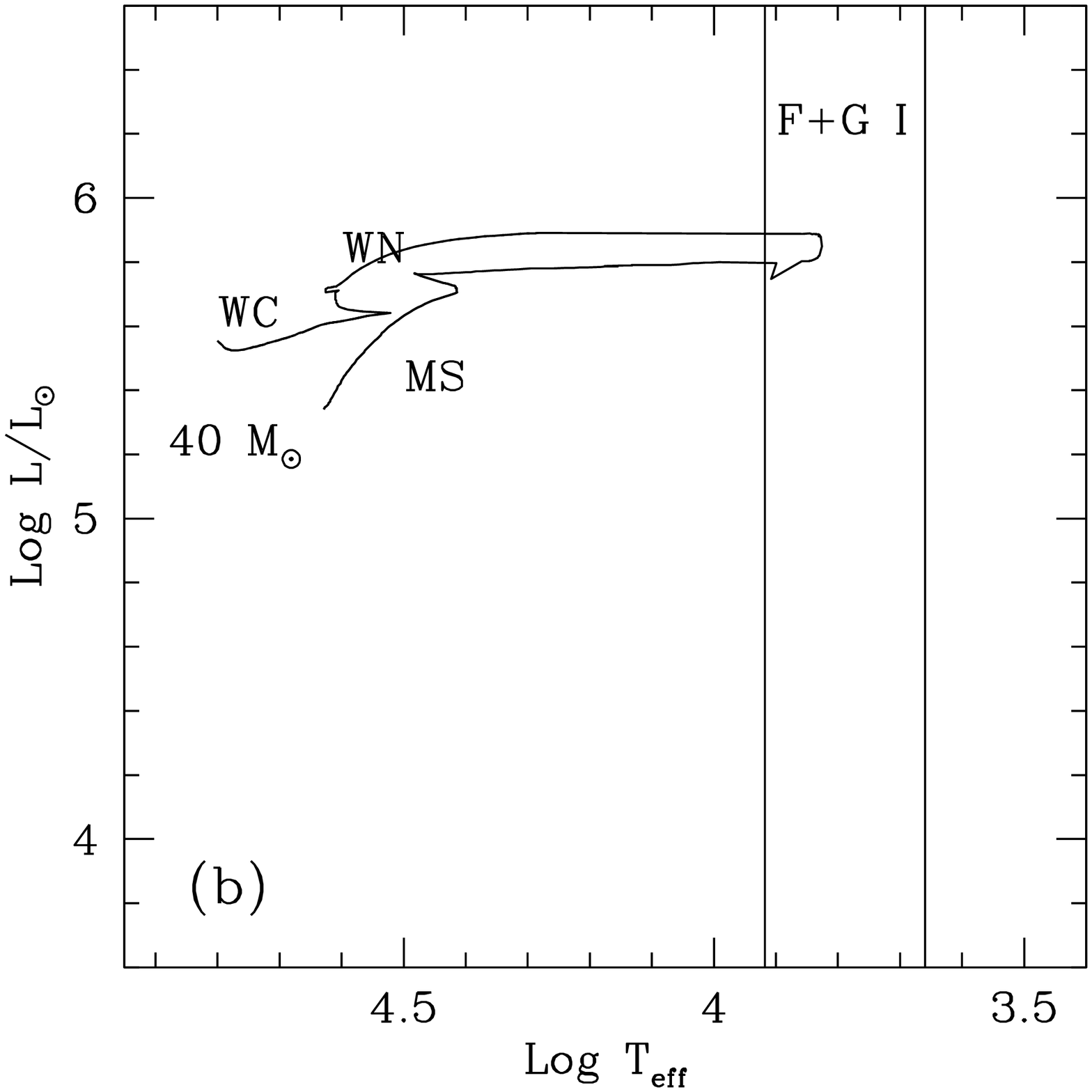}
\plottwo{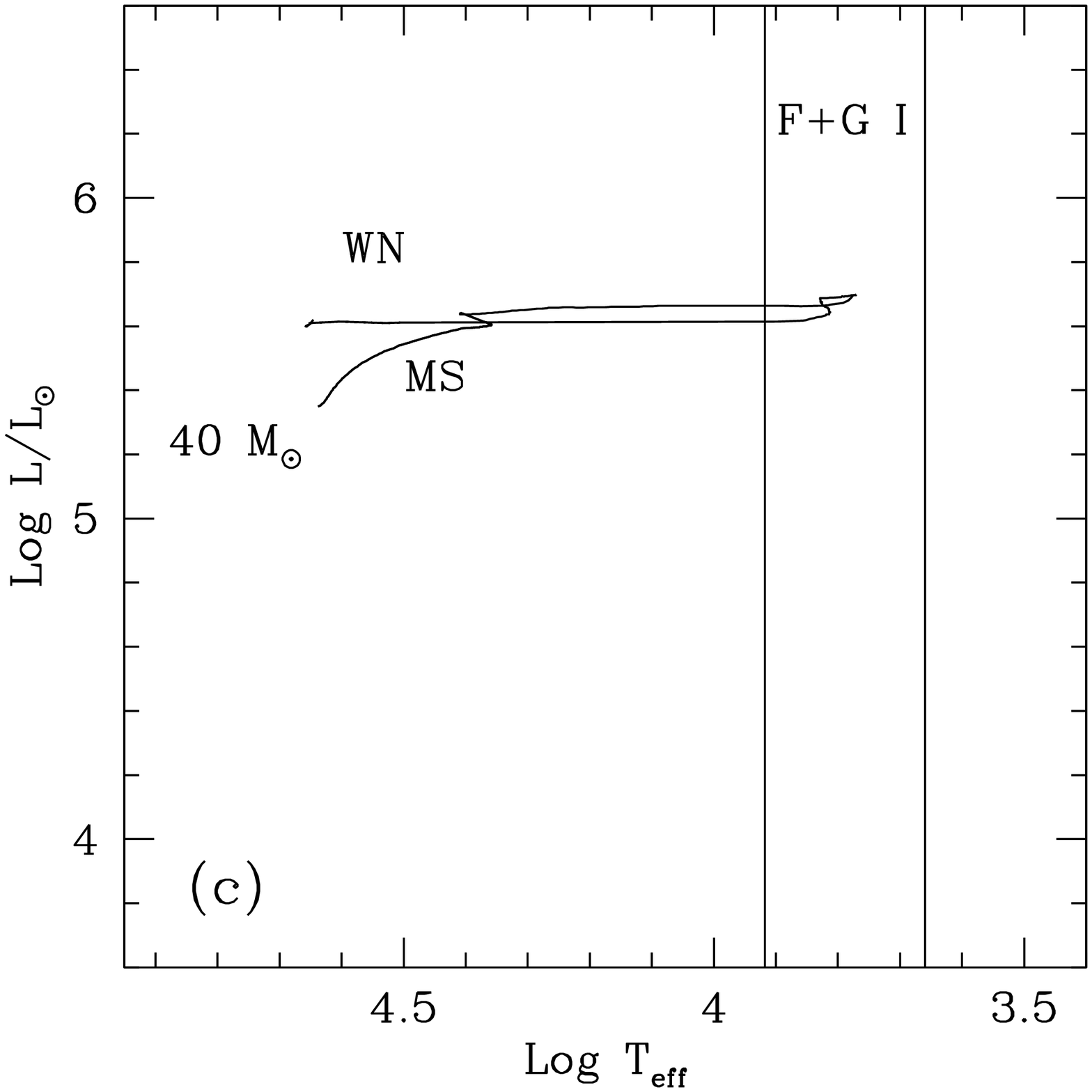}{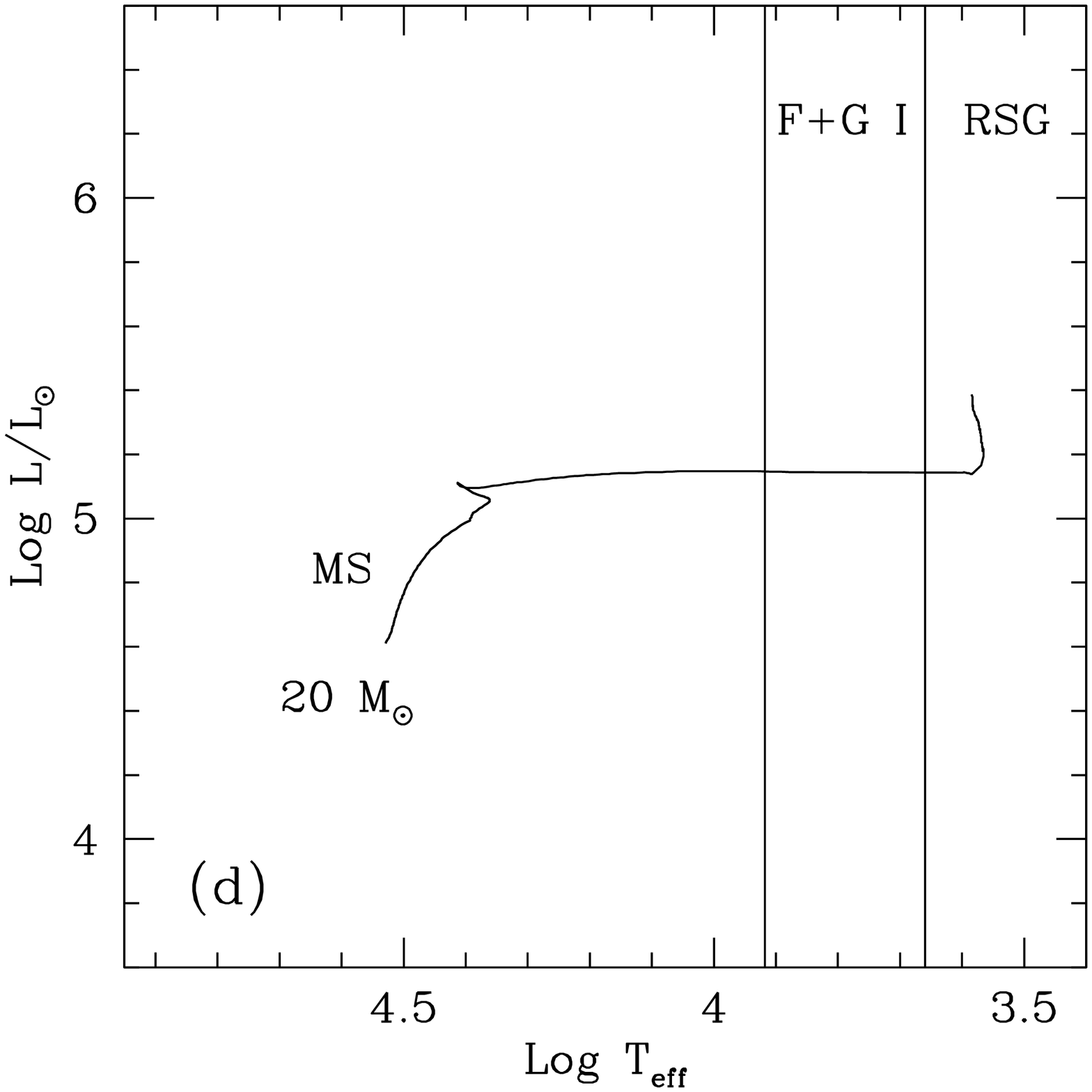}
\caption{\label{fig:HRDs} The evolution of massive stars at Galactic metallicity.  (a) The evolution of a 120$M_\odot$ star passes through a 
main-sequence phase, a Luminous Blue Variable (LBV) phase, and then through a WN- and WC-type Wolf-Rayet phase.  (b) The evolution of
a 40$M_\odot$ star passes through a main-sequence stage, a yellow (F- and G-type supergiant) phase, and then a WN- and WC-type Wolf-
Rayet stage.  (c) The evolution of a 40$M_\odot$ star never proceeds as far as the WC stage if rotation is not included. (d) The evolution of
a 20$M_\odot$ star proceeds through a yellow and then red supergiant stage.
}
\end{figure}

In Fig.~1(b) we show what the models predict will happen with a 40$M_\odot$ star: it will go through its main-sequence phase, starting out
as an O5~V star and evolving through an F and G supergiant stage.  The star then becomes hotter again, and eventually evolves to a WN-
and then a WC-type WR.  Still, making it as far as a WC is a close thing; if the calculation is done without including rotation, the star never
makes it to the WC stage, as shown in Fig.~1(c).

Finally, let us consider the evolution of a 20$M_\odot$ star.  It begins its main-sequence life as an O8~V, and then evolves through the yellow
(F and G-type) supergiant stage, finally reaching the red supergiant (RSG) stage.

\section{O-type Stars}

It would be extremely attractive to be able to use the number of unevolved massive stars with masses greater than ``X" as the normalization
factor when comparing the number of evolved massive stars of various types from one galaxy (metallicity) to another.  Indeed, it is not 
uncommon for theoreticians to do exactly that in trying to compare their model predictions to the observed number.

I don't know where these numbers come from (although sometimes my own work is cited) but historically these numbers are
 primarily a
reflection of the (in)completeness of such catalogs, and not actually the number of O stars.  The reasons are two fold.  First, the most massive
stars in a galaxy are not the brightest visually.  For instance, a 25$M_\odot$ A-type supergiant in the SMC will be found at $V=11$, while
an 85$M_\odot$ star on the ZAMS will be found at $V=14$.  {\it Some} early workers in the field recognized the limitations of their catalogs; see, for instance,
the excellent discussion of this issue in the Rousseau et al.\ (1978) catalog of LMC members.  Thus, efforts to determine the initial mass function
(IMF) of the Magellanic Clouds needed to take this into account (Massey et al.\ 1995).  The second issue is what is actually involved in
determining the luminosity of a star.  One can ``easily" obtain reddening-free photometry (with a few years of work, as we did in the Local
Group Galaxies survey), but how does one get the bolometric correction?  For hot stars most of the radiation is emitted in the far-UV, and
in the optical the dereddened spectral energy distribution is far down on the tail of the Rayleigh-Jeans distribution.  Massey (1998a) provides
an error analysis of  the derivation of masses from photometry: an uncertainty of 0.1~dex in $\log T_{\rm eff}$ results in an error of 0.15~dex in
$\log m$.  For the hottest (40,000-50,000 K) massive stars, $d\log T_{\rm eff}/dQ \sim 6$, so a modest error of 0.05~mag in the reddening-free
$Q$ color index would translate to an error in effective temperature of 0.3~dex in $\log T_{\rm eff}$, and hence about 0.45~dex in $\log m$---
in other words, you can't tell a 50$M_\odot$ star from a $150M_\odot$ star!

However, the relationships become a lot flatter for stars with effective temperatures $<$35,000 K, and it should be  possible to count 
the {\it total} number of unevolved stars.  Let us take as our limits $M_{\rm bol}<-7$ ($\log L/L_\odot>4.7$) and $T_{\rm eff}>$30,000 K,
which should roughly include all of the H-burning stars of 20$M_\odot$ and above. Since there are several ways to account for the reddening,
let's try several of them on the LGGS data for M33 and see how well the answers agree:
\begin{enumerate}
\itemsep -1pt
\item Variable reddening and determine separate transformations for each luminosity class: 22,640.
\item Adopt constant reddening: 23,450
\item Use variable reddening but treat all luminosity classes the same: 22,740.
\end{enumerate}
So, our answers are at least consistent at the 5\% level!

Although we cannot {\it directly} measure the number of massive stars above (say) 50$M_\odot$ we can now calculate the number based
on these totals, if we assume an IMF slope and main-sequence lifetime.  In general, the number of stars $n$ between mass $m_1$ and
$m_2$ will be $$n_{m_1}^{m_2}\propto [m^{\Gamma} \times \tau_{ms}(m)]_{m_1}^{m_2}$$
where $\tau_{ms}$ is the main-sequence lifetime, and $\Gamma$ is the IMF slope  (i.e., Salpeter has $\Gamma=-1.35$).  Since the
mass-luminosity relationship goes something like $L\sim m^2$ for high-mass stars, then $\tau_{ms} \sim m^{-1}$ from first principles, and
hence $n_{m_1}^{m_2}=A [m^{-2.35}]_{m_1}^{m_2}$.  We can thus use the total number of stars to find the normalization constant $A$.
The value for $m_1$ is 20$M_\odot$, and  it doesn't matter if we assume $m_2=150M_\odot$ or $\infty$.   For M33 then $A=-2.6\times10^{7}$.  The total number of stars with masses $m>50M_\odot$ is (approximately!) 2,600, and
the number of stars with masses greater than 100$M_\odot$ is 500.  If stars with masses greater than 150$M_\odot$ existed, we would expect
to find 200 of them, and for that matter, the present day mass function wouldn't peter down to one star until 1400$M_\odot$!.  Thus, one would
think that it would be
pretty straight-forward to determine the upper mass cut-off in such a mixed age population, although perhaps
greater care would be needed on things
like stellar ages.

Such an argument is flawed for the following reason.  Although some (e.g., Oey \& Clarke 2005)
 have treated the upper mass limit
from purely a statistical point of view by combining data from various clusters  to argue that in the ensemble population one should see a star of very
high mass (but one doesn't),  Kroupa \& Weidner (2003) and Weidner \& Kroupa (2005) have warned that because individual clusters
of finite mass will each have an upper boundary, one does not recover the same IMF by combining a bunch of separate groups of stars.
Thus, our estimate of even the number of stars with masses above 50$M_\odot$ is probably flawed, as there are doubtless OB associations
and clusters that contribute to the total that have upper masses less than 50$M_\odot$ (but more than 20$M_\odot$).

We include in Table 1  the total number of $20M_\odot$ stars, and (keeping in mind the above caveat) the very approximate number more massive than 50$M_\odot$.

\begin{table}[!ht]
\caption{Number of unevolved massive stars}
\smallskip
\begin{center}
{\small
\begin{tabular}{l r r r}
\tableline
\noalign{\smallskip}
Galaxy & $>20M_\odot$ & $>50M_\odot$  \\
\noalign{\smallskip}
\tableline
\noalign{\smallskip}
M31 & 24,800    & 2,800   \\
M33 & 22,600    & 2,600  \\
LMC & 6,100     & 700       \\
SMC & 2,800     & 330     \\
N6822 & 1,300  & 150      \\
WLM  &    830   & 100       \\
IC10 & 550        & 65        \\
SexA & 290        & 35       \\
Phoenix & 270   & 30        \\
SexB & 150        & 17        \\
Pegasus &130    & 15       \\
\noalign{\smallskip}
\tableline
\end{tabular}
}
\end{center}
\end{table}

\section{Red Supergiants}

Red supergiants are the He-burning descendants of 10-25$M_\odot$ stars, and mostly represent
the terminus in the life of such stars.

I first got intrigued by the issue of the red supergiant content of nearby galaxies by the
interesting study of M33 by Humphreys \& Sandage (1980).  Their Figs.\ 21 and 22
tell an interesting story: the distribution of blue stars seen against M33 are quite clumpy,
while the distribution of red stars is quite smooth.  The argument that RSGs are in general
older than O stars is quite weak as an explanation: in 10 million years a star with a random
velocity of 30 km/sec would drift by about 300 pc, or about 1.5 arcmin in M33, which is only
comparable to the size of typical OB association they identify from the blue stars---not enough
to smooth out the distribution.  The problem was eventually solved by Massey (1998b), who
found that most of the red stars identified by Humphreys \& Sandage (1980) were actually
foreground stars.  This brings us to the main issue with completing an accurate census of the
HRD for stars other than the blue ones---foreground contamination can be quite an issue.

This particular problem---separating the RSGs from the foreground stars---is one that can be
solved by photometry.  As shown in Massey (1998b), a two-color $V-R$, $B-V$ does a nice
job of separating the two.  We show such a diagram here for the LGGS data on M31.  In Fig.~2 ({\it upper}),
we see that there is a pretty clean separation.  The upper sequence are the RSGs; the lower
sequence are the (more numerous) foreground dwarfs.

\begin{figure}[!ht]
\plotfiddle{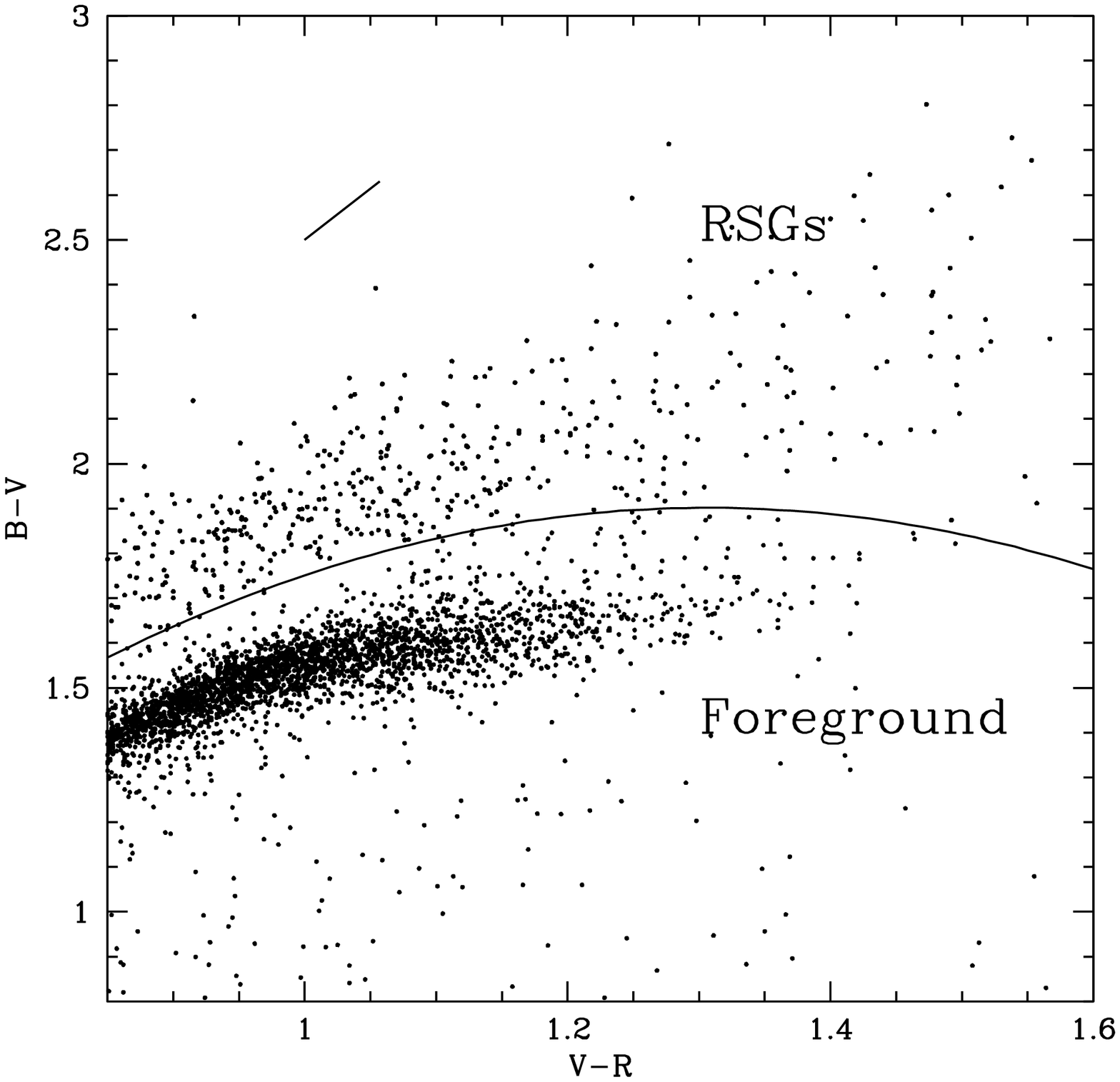}{2.75in}{0.}{35}{35}{70}{120}
\plottwo{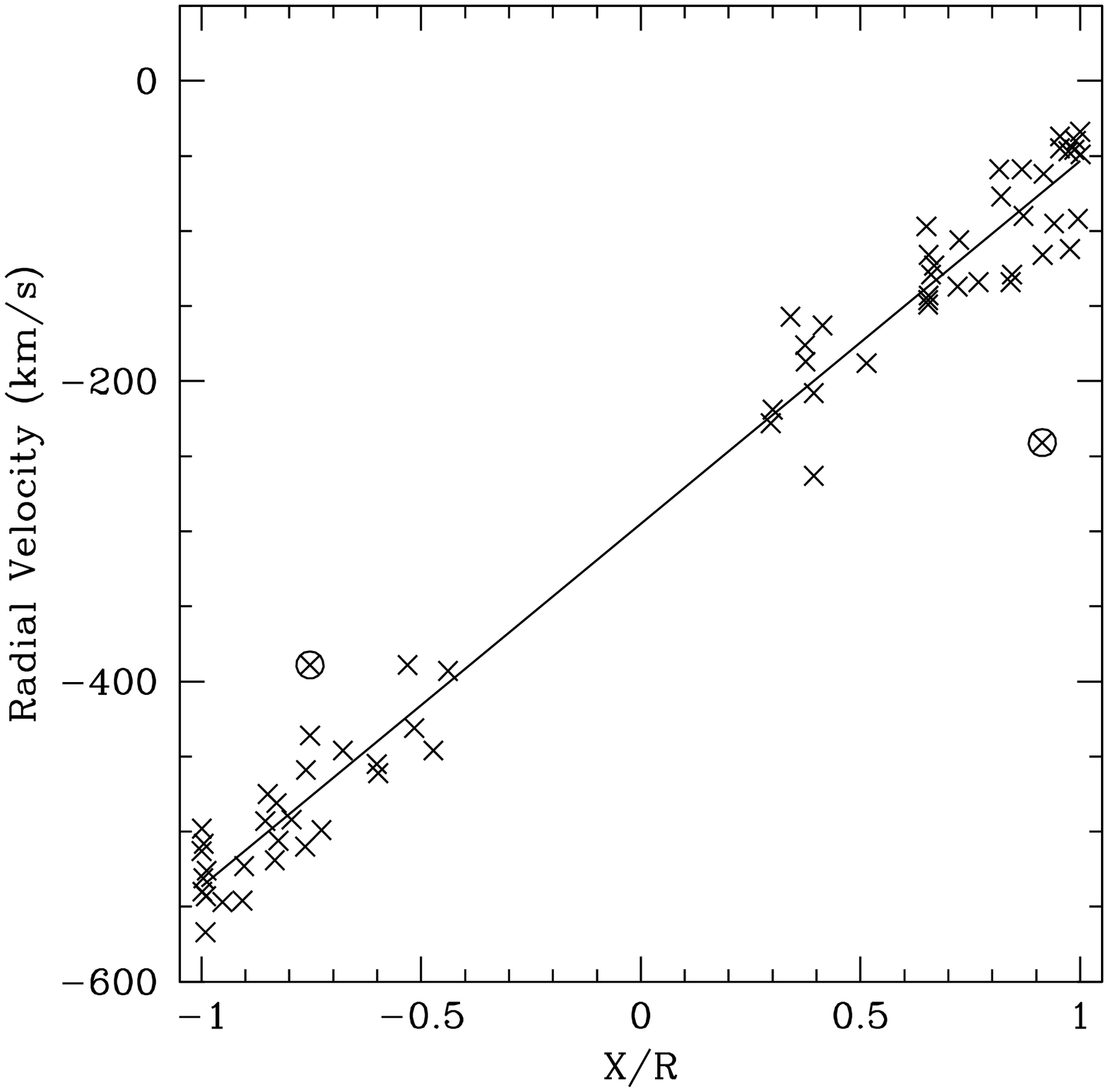}{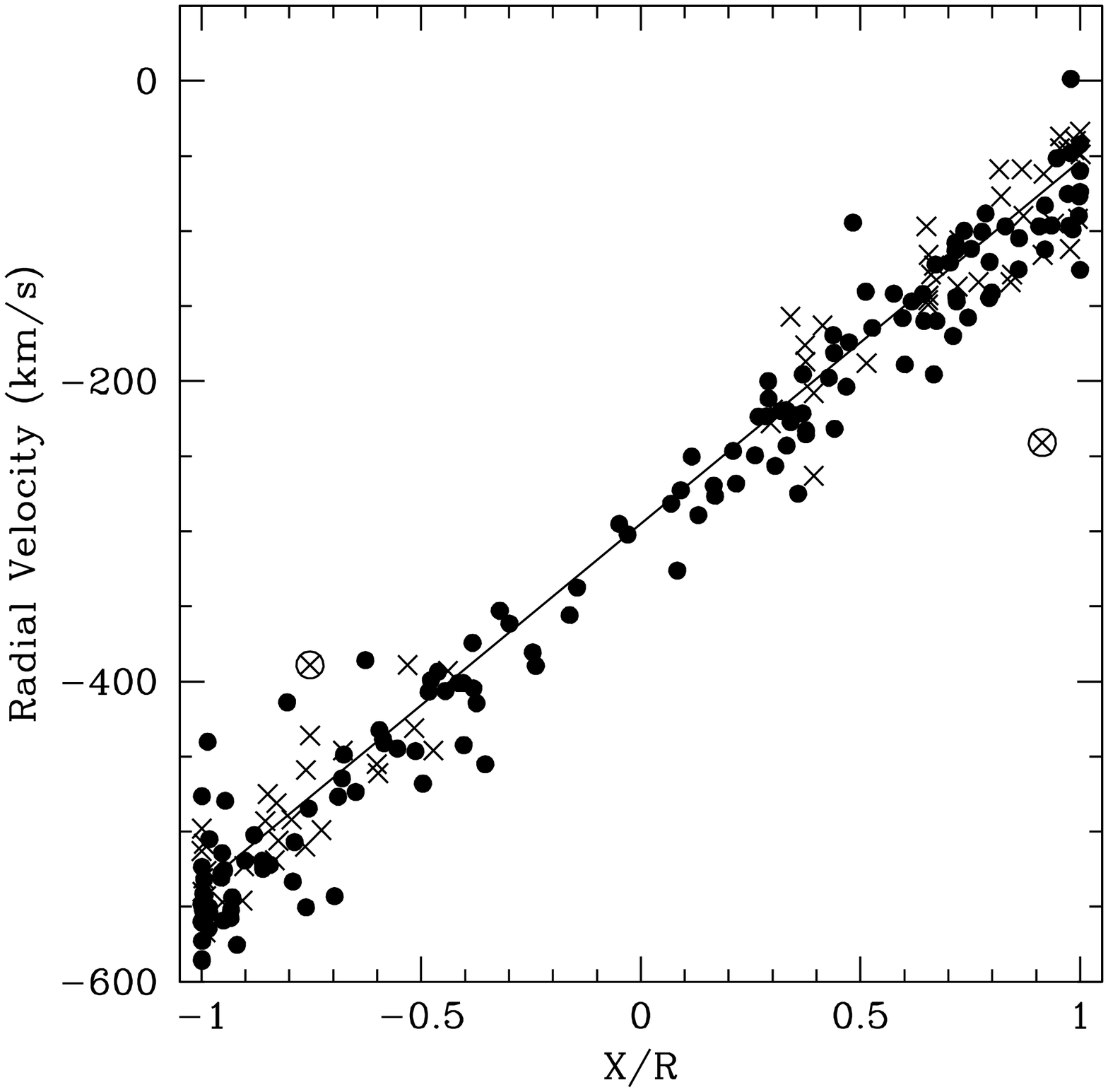}
\caption{\label{fig:2color.eps} Identifying red supergiants.  In the upper figure, we show a two-color
diagram for bright ($V<20$), red ($V-R>0.85$) stars seen against M31.  There are clearly two sequences,
which we have distinguished with a curve.  The short line at upper left is the typical reddening vector of M31.
In the lower two figures we demonstrate how well our selection criteria actually worked.  On the left is the
radial velocity of the HII regions studied by Rubin \& Ford (1970) plotted against the distance along the semi-major
axis divided by the radius.  The line is a least-squares fit to these data, excluding the two circled outliers.  On the
right we now superimpose the radial velocities of the RSG candidates selected as per the top figure. From Massey et al.\ (2009). }
\end{figure}

How well does the method work in practice?  For M31 we can test this exactly, as the galaxy has a large rotational rate
superimposed upon a large (negative) systemic velocity.  In the lower left panel of Fig.~2 we show the radial velocities of
the HII regions from the seminal study of the rotation of M31 by Rubin \& Ford (1970).  The abscissa is the distance along the
semi-major axis, normalized by the radius of the galaxy, where we have adopted the geometrical parameters of Rubin \& Ford (1970).
The fact that a straight line fits these data is a dramatic demonstration that the gravitational potential of M31 is dominated by dark
matter, in that the rotation curve is {\it really} flat.  On the right we now superimpose the radial velocities of the RSGs selected on the
basis of the two-color diagram.  The agreement says it all: it {\it worked!}   This argument comes
from the study of RSGs in M31 by Massey et al.\ (2009).

So what are the difficulties here?  Well, there are actually only two caveats, both relatively minor: first, for a meaningful comparison, we really
need to cut at a $\log L/L_\odot>4.5$, else we run the risk of including intermediate-mass asymptotic giant branch stars in our census.  This
potential issue was first noted by Brunish et al.\ (1986).  The second issue, and philosophically more subtle, is what exactly do we mean
by a ``red" supergiant?  From a stellar evolutionary point of view, red supergiants are those things
over on the right side of the HRD which blow up as spectacular supernovae, right?
 Well, not from an observer's point of view.  The difficulty comes in that a 25$M_\odot$
``red supergiant" at SMC metallicities actually has an effective temperature of about 5550~K (Charbonnel et al.\ 1993).  In other words,
the terminus of the star's He-burning phase occurs as an G supergiant, not a K-M~I.  So, when we compare our census to the expectations
of stellar evolutionary theory, we have to make sure that we all agree on what is what.

\section{The Yellow Supergiants}

As we showed  in \S~2 (Fig.~1), the yellow supergiant (F and G~I) will be the ``turning around" point in the evolution of a
40$M_\odot$ star, while stars of lower masses will briefly pass through a yellow supergiant phase on their way to becoming red supergiants.
The yellow supergiant phase is a very appealing one to study as the lifetimes are extremely short (2,500-500,000 years).  However, it
is also a very difficult phase to study for two reasons: (a) the foreground contamination is very large (90-95\% for most Local Group galaxies),
and (b) there are no good two-color discriminants that allow one to separate out foreground from bona fide extragalactic supergiants,
unlike the situation for RSGs.

Maria Drout and I spent last summer working on this problem, again using the rotation of M31 to separate the wheat from chaff. (More details can be found in her contribution in this conference.)  We began by
using the LGGS photometry to select 5000 stars whose color/magnitude indicated that they {\it could} be yellow supergiants.  Using the
mighty Hectospect fiber spectrometer on the 6.5-m MMT we obtained spectra of $\sim$ 3000 of these stars so we could measure their
radial velocities.  In Fig.~3 (left) I show the observed velocities (dashed line) compared to those expected from foreground alone (solid line),
where the latter is based on the Besancon model (Robin et al.\ 2003).
Of these 3000 stars, 54 of them  turn out to be ``definite" supergiants based upon their radial velocities.  For
another 66 we can't tell for sure, as their expected velocities are not that different from a halo star.   The resulting H-R diagram
is shown in Fig.~3 (right).  The surprising thing we find is that nearly {\it all} of the yellow supergiants have masses of 20$M_\odot$ and less.
There are very few even ``possible" supergiants above this.  Yet, the evolutionary tracks predict longer 
(or comparable) life times for the higher
mass yellow supergiants, and thus---even given the exponential nature of the IMF---we would expect to find as many higher mass stars as lower.  The discrepancy is large, and suggests to us
that something is wrong with the lifetimes predicted by the evolutionary tracks.

\begin{figure}[!ht]
\plottwo{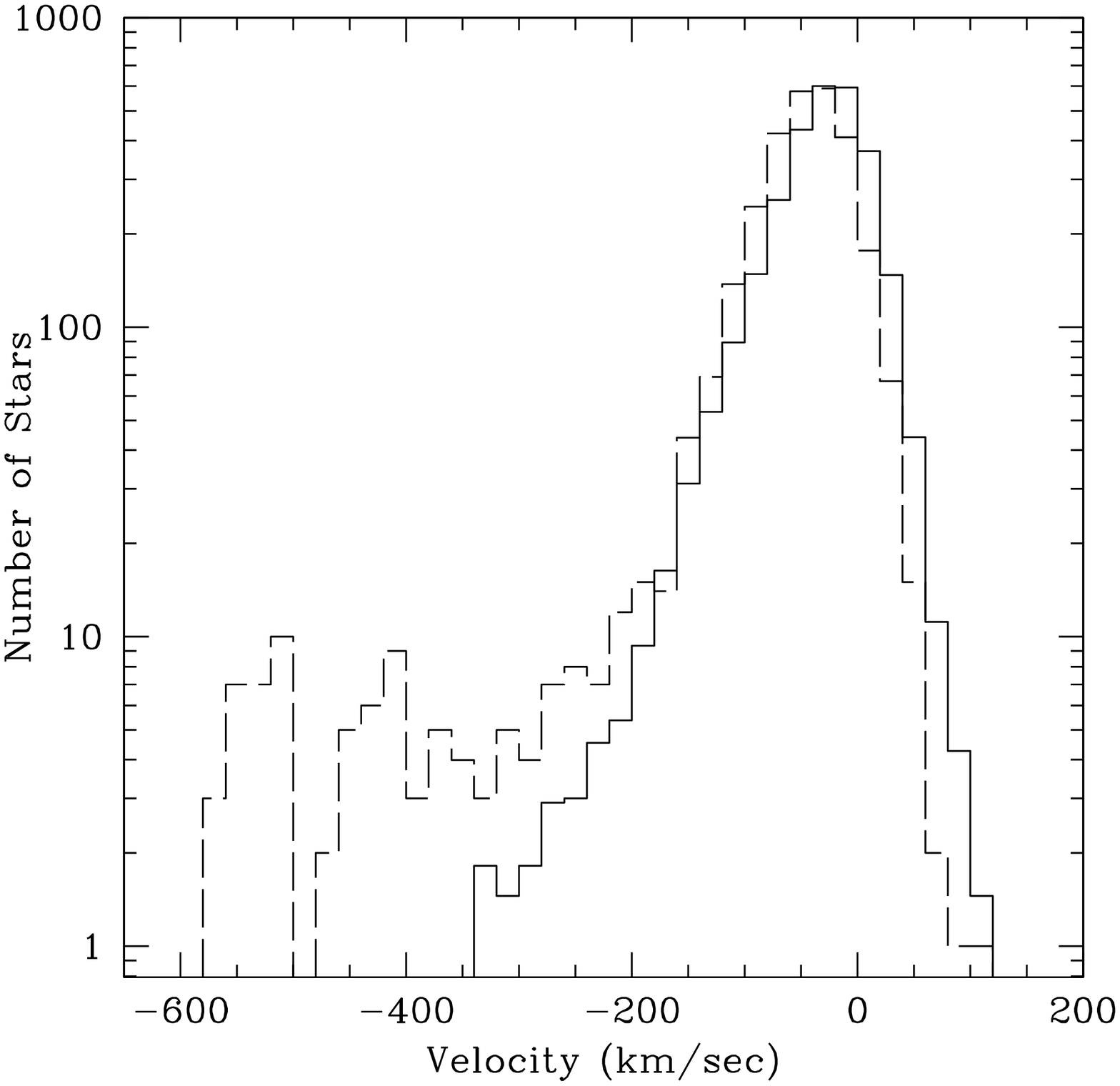}{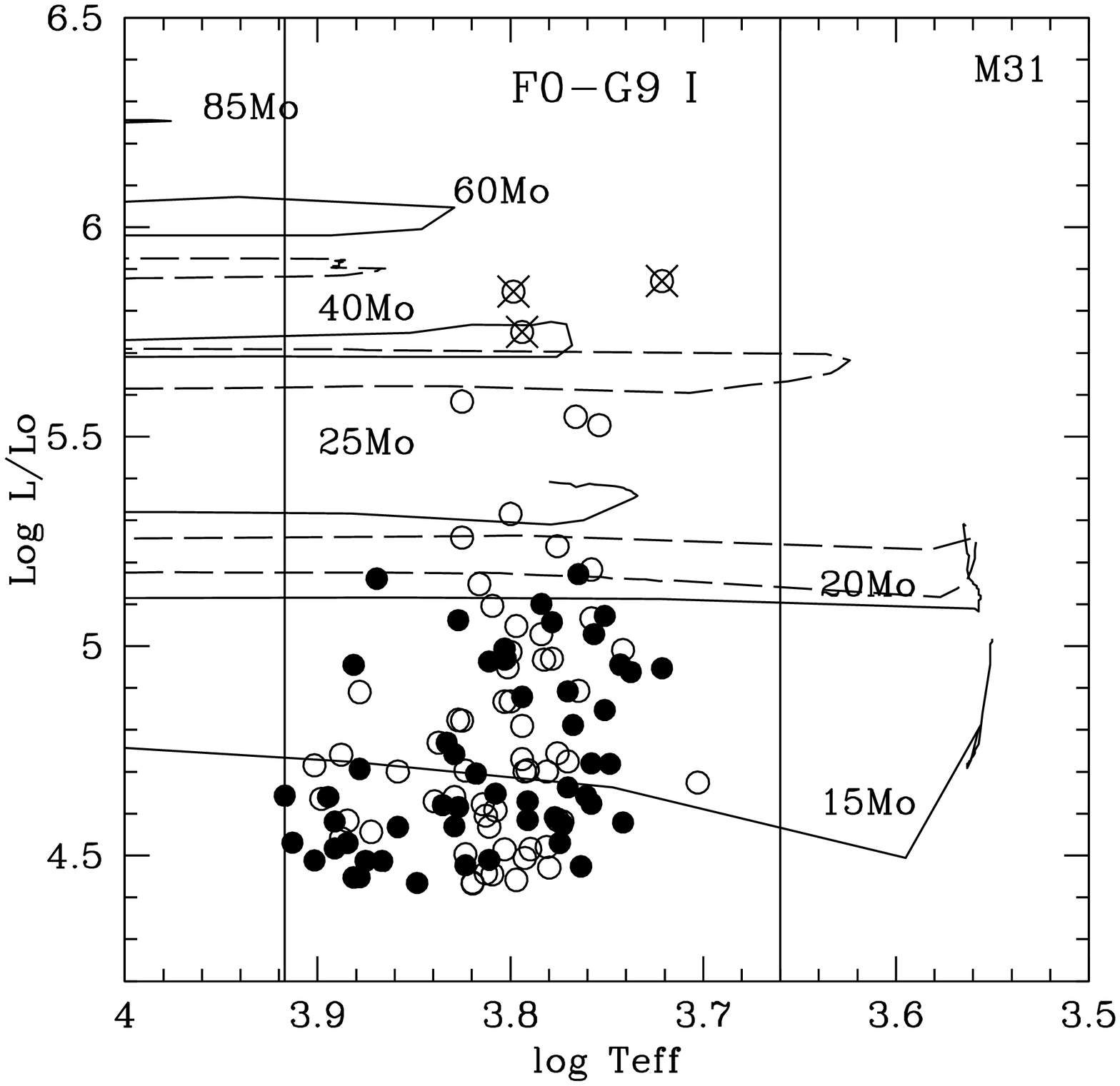}
\caption{\label{fig:fstars.eps} Yellow supergiants in M31.  On the left, the dashed line shows the observed radial velocities of yellow stars, while the solid line denotes that expected for foreground stars alone.
The excess at negative velocities contains the bona fide yellow supergiants.  
On the right we see the H-R diagram for the yellow supergiants.  Solid points are
the ``certain" members, while the open circles denote the ``possible" members.  The three brightest such stars, with x's through them,
have now been ruled out as yellow supergiants based upon detailed spectroscopy of luminosity sensitive features.}  
\end{figure}

\section{Special Cases}

Here I will briefly discuss the two ``special cases", that of detecting Wolf-Rayet stars (WRs) and Luminous Blue Variables (LBVs).

\subsection{Wolf-Rayet Stars}

The relative number of WC- and WN-type WRs has long been known to vary among the galaxies of the Local Group, being (roughly)
1:1 in the solar neighborhood, 1:5 in the LMC, and 1:11 in the SMC (see Massey \& Johnson 1998 and  Massey et al.\ 2003).  Furthermore,
M33 shows a galactro-centric gradient in the WC/WN ratio (Massey \& Conti 1983).  These data suggest that there is a simple 
correlation between the WC to WN ratio and the metallicity of a region, which is in accord with our basic understanding
of stellar evolutionary theory, i.e., at high metallicities it is easier to produce a WC-type WR as the mass-loss rates are higher 
and hence it is easier to ``peel" down the outer layers to reveal the products of He-burning at the surface.

 The difficulties of finding WRs, and the selection effects involved, have been discussed in detail by Massey \& Johnson (1998),
and I will only briefly summarize the current situation.  First, the most efficient way to find WRs is by on-line, off-line interference
filter imaging, with follow-up spectroscopy used to check the marginal cases.  The difficulty is that WC-type WRs have much stronger
emission lines than do WNs, and hence are much easier to find.  Limited regions have been deeply surveyed in M31, M33, and
NGC~6822 (Massey \& Johnson 1998 and references therein), as well as the SMC (Massey \& Duffy 2001).  Most, but probably
not all, of the WRs in the LMC have been found (Breysacher et al.\ 1999).  The situation is a little more dicey for IC10.  Because of
the very high reddening we at one time {\it thought} we found most of the WRs (Massey \& Armandroff 1995).  The number was 
unexpectedly large, given how small the galaxy is, and WC types dominated, despite the fact that the metallicity was low.
However, a follow-up study by Massey \& Holmes (2002) suggested that the previous number had been badly underestimated,
and that the WC/WN ratio might be normal.  Of all of the galaxies in the Local Group, though, the statistics for the solar neighborhood
are probably the least complete, given the difficulties of variable reddening; see the discussion in Massey \& Johnson (1998).  The current
evolutionary models that include rotation however do not do a very good job of reproducing the observed number, as noted
by Meynet \& Maeder (2005), and shown here in Fig.~4.

There is clearly more work that can be done here: we have in hand deeper data, with complete areal coverage on M31 and M33, and
additional spectroscopic follow-up of the Massey \& Holmes WR candidates in IC 10 are needed.

\begin{figure}[!ht]
\plotfiddle{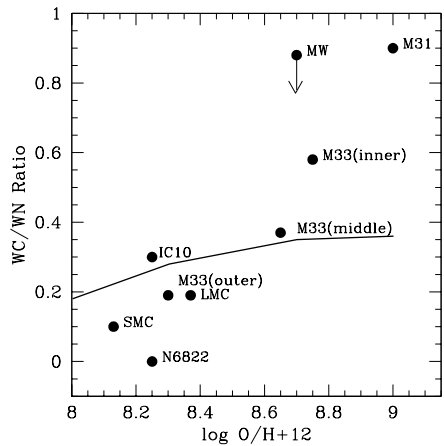}{2.0in}{0.}{30}{30}{-100}{-50}
\caption{\label{fig:wcwn.eps} The WC/WN as a function of metallicity.  The solid line denotes the theoretical prediction from
Meynet \& Maeder (2005).}
\end{figure}

\subsection{LBVs:}

LBVs are a rare class of luminous stars which undergo episodic mass loss.  As such, they likely represent a transitional phase beteen
the most masssive O stars and the WR stage (see Fig.~1).  Nebulae around the archetype LBVs $\eta$ Car and P Cyg reveal
very high ejecta masses, of order 10$M_\odot$, and evidence of multiple ejections on the timescales of $10^3$ years.  Massey (2003)
has argued that if $\eta$ Car or P Cyg were located in M31 or M33 we would not know of them today, since their spectacular photometric outbursts 
were hundreds of years ago.    Indeed, Hubble \& Sandage (1953) discovered only five such 
variables in all of M31 and M33 using photographic records that went back 40 years.  

Let me argue that if anything P Cyg might make a better poster child for LBVs than the (binary) $\eta$ Car.  The last time it did anything photometrically
interesting was in 1655.  If a similar star was in M31, how would you recognize it?  Well, you might blunder across it spectroscopically, as I happened
to one evening: the LGGS star J004341.84+411112.0 was spectroscopically {\it indistinguishable} from that of P Cyg, and the similarities didn't
end there (Massey 2006).  In particular, the M31 star showed extension of an {\it HST} image that was suggestive of a previous outburst 2000
years earlier.

We went looking for stars with spectra similar to the known LBVs, using as our candidate list stars picked our from the interference filter
imaging of the LGGS.   And did we ever find them!  Rather than being able to count the M31 and M33 LBVs on the fingers of one hand, we
believe the actual numbers are in the hundreds (Massey et al.\ 2007b).   Are these ``actual" LBVs?  If flashy 
photometric variability is a pre-requeste for membership in the LBV club, we may have to wait a long time to be sure---say, a millennium or two.
Or, we could use {\it HST} to detect signs of past ejecta events...if we can ever convince the TAC!


\acknowledgements 

Much of the work described herein has relied upon collaborations with a number of colleagues, including
Nelson Caldwell,
Geoff Clayton,
Maria Drout,
Paul Hodge,
George Jacoby,
Emily Levesque,
Andre Maeder,
Taylor McNeill,
Georges Meynet,
Knut Olsen,
Bertrand Plez,
Dave Silva, 
and
Susan Tokarz.
In addition, Deidre Hunter and Emily Levesque made useful comments on an early draft of this manuscript.
This work was supported by the National Science Foundation through AST-0604569 and AST-0844315.



\begin{thebibliography}{}

\bibitem[Brunish et al.\ (1986)]{Brunish86}
Brunish, W. M., Gallagher, J. S., \& Truran, J. W. 1986, AJ, 91, 598

\bibitem[]{} Breysacher, J., Azzopardi, M., \& Testor, G. 1999, A\&AS, 137, 117

\bibitem[]{}
Charbonnell, C., Meynet, G., Maeder, A, Schaller, G., \& Schaerer, D. 1993,
A\&AS, 101, 415

\bibitem[Conti (1976)]{Conti76}
Conti,  P. S. 1976, Mem.~Soc.~R.~Sci.~Liege, 
9, 193

\bibitem[]{} 
Hubble, E., \& Sandage, A. 1953, ApJ, 118, 353

\bibitem[Humphreys \& Sandage (1980)]{HumpSand80}
Humphreys, R. M., \& Sandage, A. 1980, ApJS, 44, 319

\bibitem[Kroupa \& Weidner (2003)]{KroupaWeidner03}
Kroupa, P., \& Weidner, C. 2003, ApJ, 598, 1076


\bibitem[Massey(1998a)]{Massey98IMF}
Massey, P. 1998a, in The Stellar Initial Mass Function, 
38th Herstmonceux Conference, ed. G. Gilmore \& D. Howell
(San Francisco: ASP),  17

\bibitem[Massey(1998b)]{Massey98RSG}
Massey, P. 1998b, ApJ, 501, 153

\bibitem[Massey(2003)]{Massey03}
Massey, P. 2003, ARA\&A, 41, 15

\bibitem[]{} Massey, P. 2006, ApJ, 638, L93

\bibitem[]{} Massey, P., \& Armandroff, T. E. 1995, AJ, 109, 2470

\bibitem[]{} Massey, P., \& Conti, P. S. 1983, ApJ, 273, 576

\bibitem[]{} Massey, P., \& Duffy, A. S. 2001, ApJ, 550, 713

\bibitem[Massey \& Johnson(1998)]{MasseyJohnson98}
Massey,  P. \&  Johnson,  O. 1998
{\it Ap.~J.}
505, 793

\bibitem[]{} Massey, P., \& Holmes, S. 2002, ApJ, 580, L35

\bibitem[]{} Massey, P., Olsen, K. A. G., \& Parker, J. W. 2003, PASP, 115, 1265

\bibitem[Massey et al.\ (1995)]{IMF95}
Massey, P., Lang, C. C., DeGioia-Eastwood, K., \& Garmany, C. D. 1995, ApJ, 438, 188

\bibitem[Massey et al.\ (2006)]{LGGSI}
Massey, P. et al.\  2006, AJ, 131, 2478

\bibitem[Massey et al.\ (2007a)]{LGGSII}
Massey, P. et al.\  2007a, AJ, 133, 2393

\bibitem[Massey et al.\ (2007b)]{LGGSIII}
Massey, P. et al.\  2007b,
AJ, 134, 2474

\bibitem[Massey et al.\ (2009)]{M31RSGs}
Massey, P. et al.\  2009, ApJ, submitted


\bibitem[Meynet \& Maeder(2003)]{PaperX}
Meynet, G., \& Maeder, A. 2003, A\&A, 404, 543

\bibitem[Meynet \& Maeder(2005)]{PaperXI}
Meynet, G., \& Maeder, A. 2005, A\&A, 429, 581

\bibitem[Oey \& Clarke(2005)]{OeyClarke05}
Oey, M. S., \& Clarke, C. J. 2005, ApJ, 620, L43

\bibitem[]{} Robin, A. C., Reyle, C., Derriere, S., \& Picaud, S. 2003, A\&A, 509, 523

\bibitem[Rousseau et al.\ (1978)]{Rou78}
Rousseau, J., Martin, N., Prevot, L., Rebeirot, E., Robin, A., \& Brunet, J. P. 1978, A\&AS, 31, 243

\bibitem[Rubin \& Ford (1970)]{RubinFord70}
Rubin,  V. C., \& Ford, W. K. J. 1970, ApJ, 159, 379

\bibitem[Vanbeveren \& Conti(1980)]{VanbevConti80}
Vanbeveren, D., \& Conti, P. S. 1980, A\&A, 88, 230


\bibitem[Vink et al.(2001)]{Vinketal01}
Vink, J. S., de Koter, A., Lamers, H. J. G. L. M. 2001, A\&A, 369, 574

\bibitem[Weidner \& Kroupa (2005)]{WK05}
Weidner, C., \& Kroupa, P. 2005, ApJ, 625, 754


\end{thebibliography}
\end{document}